\documentclass[letterpaper]{article} % DO NOT CHANGE THIS
\usepackage{aaai24}  % DO NOT CHANGE THIS
\usepackage{times}  % DO NOT CHANGE THIS
\usepackage{helvet}  % DO NOT CHANGE THIS
\usepackage{courier}  % DO NOT CHANGE THIS
\usepackage[hyphens]{url}  % DO NOT CHANGE THIS
\usepackage{graphicx} % DO NOT CHANGE THIS
\usepackage{natbib}  % DO NOT CHANGE THIS AND DO NOT ADD ANY OPTIONS TO IT
\usepackage{caption} % DO NOT CHANGE THIS AND DO NOT ADD ANY OPTIONS TO IT
\usepackage{algorithm}
\usepackage{algorithmic}
\usepackage{newtxmath} 
\usepackage{booktabs}
\usepackage{multirow}
\usepackage{makecell}
\usepackage{newfloat}
\usepackage{listings}
\DeclareCaptionStyle{ruled}{labelfont=normalfont,labelsep=colon,strut=off} % DO NOT CHANGE THIS
\floatstyle{ruled}
\newfloat{listing}{tb}{lst}{}
\floatname{listing}{Listing}
\title{T2MAC: Targeted and Trusted Multi-Agent Communication through Selective Engagement and Evidence-Driven Integration}
\author{
    Chuxiong Sun\textsuperscript{\rm 1 \rm 2}\equalcontrib,
    Zehua Zang\textsuperscript{\rm 1 \rm 3}\equalcontrib,
    Jiabao Li\textsuperscript{\rm 4}\equalcontrib,
    Jiangmeng Li\textsuperscript{\rm 1 \rm 2}\thanks{Corresponding author.},
    Xiao Xu\textsuperscript{\rm 2},
    Rui Wang\textsuperscript{\rm 1 \rm 2 \rm 3},
    Changwen Zheng\textsuperscript{\rm 1 \rm 3}
}
\affiliations{
    \textsuperscript{\rm 1}Science \& Technology on Integrated Information System Laboratory, Institute of Software Chinese Academy of Sciences\\
    \textsuperscript{\rm 2}State Key Laboratory of Intelligent Game\\
    \textsuperscript{\rm 3}University of Chinese Academy of Sciences\\
    \textsuperscript{\rm 4}School of Automation and Electrical Engineering, University of Science and Technology Beijing\\

    \{chuxiong2016,zehua2020\}@iscas.ac.cn, M202110548@xs.ustb.edu.cn, jiangmeng2019@iscas.ac.cn, xuxiao0825@gmail.com, \{wangrui,changwen\}@iscas.ac.cn,
}

\usepackage{bibentry}
\begin{document}

\maketitle

\begin{abstract}
Communication stands as a potent mechanism to harmonize the behaviors of multiple agents. 
However, existing works primarily concentrate on broadcast communication, which not only lacks practicality, but also leads to information redundancy. 
This surplus, one-fits-all information could adversely impact the communication efficiency. 
Furthermore, existing works often resort to basic mechanisms to integrate observed and received information, impairing the learning process. 
To tackle these difficulties, we propose Targeted and Trusted Multi-Agent Communication (T2MAC), a straightforward yet effective method that enables agents to learn selective engagement and evidence-driven integration.  
With T2MAC, agents have the capability to craft individualized messages, pinpoint ideal communication windows, and engage with reliable partners, thereby refining communication efficiency. 
Following the reception of messages, the agents integrate information observed and received from different sources at an evidence level. This process enables agents to collectively use evidence garnered from multiple perspectives, fostering trusted and cooperative behaviors. 
We evaluate our method on a diverse set of cooperative multi-agent tasks, with varying difficulties, involving different scales and ranging from Hallway, MPE to SMAC. The experiments indicate that the proposed model not only surpasses the state-of-the-art methods in terms of cooperative performance and communication efficiency, but also exhibits impressive generalization.

\end{abstract}

\section{Introduction}
% why marl -> why communication? 1.partial observation 2. non-stationary
Reinforcement Learning (RL) has achieved remarkable milestones in a myriad of intricate real-world domains, ranging from Game AI \cite{osband2016deep,silver2017mastering,silver2018general,vinyals2019grandmaster} and Robotics \cite{robotics} to Autonomous Driving \cite{carla}. However, when delving into cooperative multi-agent settings, distinct challenges surface. The issue of partial observability stands out, where agents are confined to their local observations, missing out on the broader perspective of the entire environment. Complicating matters further, Multi-Agent Reinforcement Learning (MARL) grapples with the non-stationarity of the environment. From an individual agent's perspective, the environmental dynamics shift incessantly, adding another layer of complexity to the learning process.

Multi-agent communication offers a compelling solution to the issues outlined by granting agents the capability to derive a deeper understanding of their surroundings through collective insights. 
This approach ensures stable learning and encourages harmonized actions among agents.
However, historical methods have focused on the content and timing of communication \cite{commnet111,ic3net,schednet,ndq,tmc,maic}. Once an agent elects to share its message, it is broadcast to the entire agent group. This indiscriminate broadcasting is not only resource-intensive but also potentially inefficient. 

A pivotal realization is that only some agents carry valuable insights, and flooding the network with redundant information can be counterproductive to learning.
%It's worth noting that humans possess an innate ability to discern with whom they should communicate and generate tailored messages for specific partner. By drawing inspiration from such human tendencies, we can enhance the efficiency of information exchange, equipping agents with the capability to selectively identify their communication messages and counterparts.
%targeted communication
Interestingly, humans know when to communicate intrinsically, with whom, and how to customize their messages to the recipient. Mirroring these human instincts could significantly refine the information exchange process, allowing agents to curate their messages and recipients selectively.

Moreover, the essence of messages relayed by agents is a distillation of their individual observational experiences. 
Assimilating these messages aptly can enrich agents' perception of an uncertain environment, leading to more refined policies.  
Regrettably, the existing techniques—whether they're steeped in basic aggregation \cite{jiang2018learning} or are more avant-garde with representation learning \cite{tarmac,masia} tend to treat the fusion of information as a black box, presupposing that the policy networks can innately sift out vital data and diminish decision-making uncertainty. In this context, the information integration process might prove to be both uncertain and inefficient, especially in intricate scenarios. As such, there's a pressing need for a novel and theory-grounded approach that can adeptly merge messages while tackling the  inherent underlying uncertainties. 

With this vision, we introduce the Targeted and Trusted Multi-Agent Communication (T2MAC) framework, 
which embodies the principles of discerning and streamlined communication, drawing inspiration from human inclinations to engage selectively with trusted and relevant counterparts, ensuring more efficient information integration, and fostering a more adaptive multi-agent collaboration in dynamic environments. 
Specifically, each agent is skilled at analyzing observations to extract evidence. In this context, evidence denotes metrics instrumental in guiding the decision-making processes. 
This evidence plays a dual role: guiding local decision-making and serving as the basis for crafting messages that are meticulously tailored to specific agent contexts. 
Moreover, we evaluate the variations in uncertainty prior to and post-communication to measure the impact and significance of specific communication behavior. 
Armed with these insights, we craft binary pseudo-labels based on the significance of communication and devise an auxiliary task. This task is specifically designed to train a communication selector network, empowering it to identify the ideal communication counterparts. By adopting this strategy, we guarantee that only the most relevant and credible data is exchanged among the agents. 
Upon receipt, messages are integrated at the evidence level rather than the conventional observation or feature level. 
To capture the intricacies of decision-making, we leverage the Dirichlet distribution. This allows us to model decision policies, anchoring them on evidence that's been sourced from a myriad of perspectives. 
Concretely, we integrate Subjective Logic (SL) \cite{sl} to link the Dirichlet parameters with belief and uncertainty, therefore quantifying the uncertainty for decision-making and jointly modeling the probability of each action. 
Then, we utilize Dempster-Shafer theory of evidence (DST) \cite{dst} to integrate evidence observed from multiple agents, producing a comprehensive belief and uncertainty that considers all available evidence, ensuring trusted message integration and decision-making. 
We subjected T2MAC to rigorous testing across various MARL environments, such as Hallway, MPE, and SMAC. Compared to prominent multi-agent communication strategies like TarMac \cite{tarmac}, MAIC \cite{maic}, SMS \cite{sms}, and MASIA \cite{masia}, T2MAC consistently excelled in both performance and efficiency. Additionally, its versatility shone through across diverse scenarios. 
%In conclusion, we design a multi-agent communication method, grounded in theory, 

\section{Related Works}
%decpomdp -> communication paradigm h_i^t, g_i^t, c_i^t
MARL has undergone remarkable progression in recent epochs \cite{maddpg,vdn,qmix,mappo}. Within the MARL ambit, multi-agent communication has emerged as an indispensable aspect, particularly salient for cooperative endeavors constrained by partial observability. Research in this domain can be broadly segmented into three main categories.

\textit{Deciding What to Communicate.} Historically, communication vocabularies are set in stone during training, as illustrated by \cite{foerster2016learning}. This seemingly efficient strategy unintentionally limits the depth and flexibility of agent communication.
In response, CommNet \cite{commnet111} introduces a paradigm shift by allowing agents to create dynamic, continuous messages. With its design for continuous interactions, CommNet ensures that messages are timely and sensitive to environmental changes. Building on this foundation, both VBC \cite{vbc} and TMC \cite{tmc} further optimize message learning processes. Furthermore, NDQ \cite{ndq} and MAIC \cite{maic} are designed to craft messages tailored for individual agents.

\textit{Deciding When and With Whom to Communicate.} Effective communication timing and partner selection are pivotal in Multi-Agent Communication. A gating network showcased in \cite{ic3net,jiang2018learning} generates binary decisions, allowing agents the freedom to communicate or abstain. Advancing this idea, \cite{schednet,mao2019learning,wang2020bottleneck,immac} implement a weight-based scheduler, prioritizing agents holding vital observations. Enriching this approach, I2C \cite{I2C1}, MAGIC \cite{magic1}, and SMS \cite{sms} harness methods like causal inference, graph-attention, and Shapley message value to pinpoint ideal communication recipients.

\textit{Incorporating Messages for Cooperative Decision-Making.} A prominent subset of the MARL methodologies posits an egalitarian weightage to all incoming messages. Such an approach fails to recognize the significance of filtering vital information from a sea of communications. Therefore, we introduce representation learning paradigms to address this lacuna for discerning message assimilation. For instance, TarMac \cite{tarmac} adopts soft attention mechanisms to weight messages, while MASIA \cite{masia} consolidates received messages into concise representations using an autoencoder.

To our knowledge, no existing MARL method simultaneously addresses targeted and trusted communication. T2MAC stands as the pioneering approach, enabling agents to efficiently select communication partners and distill tailored evidence and integrate messages, resulting in trustworthy cooperative decisions.

\begin{figure*}[t]
  \centering
  \includegraphics[width=1.0\linewidth]{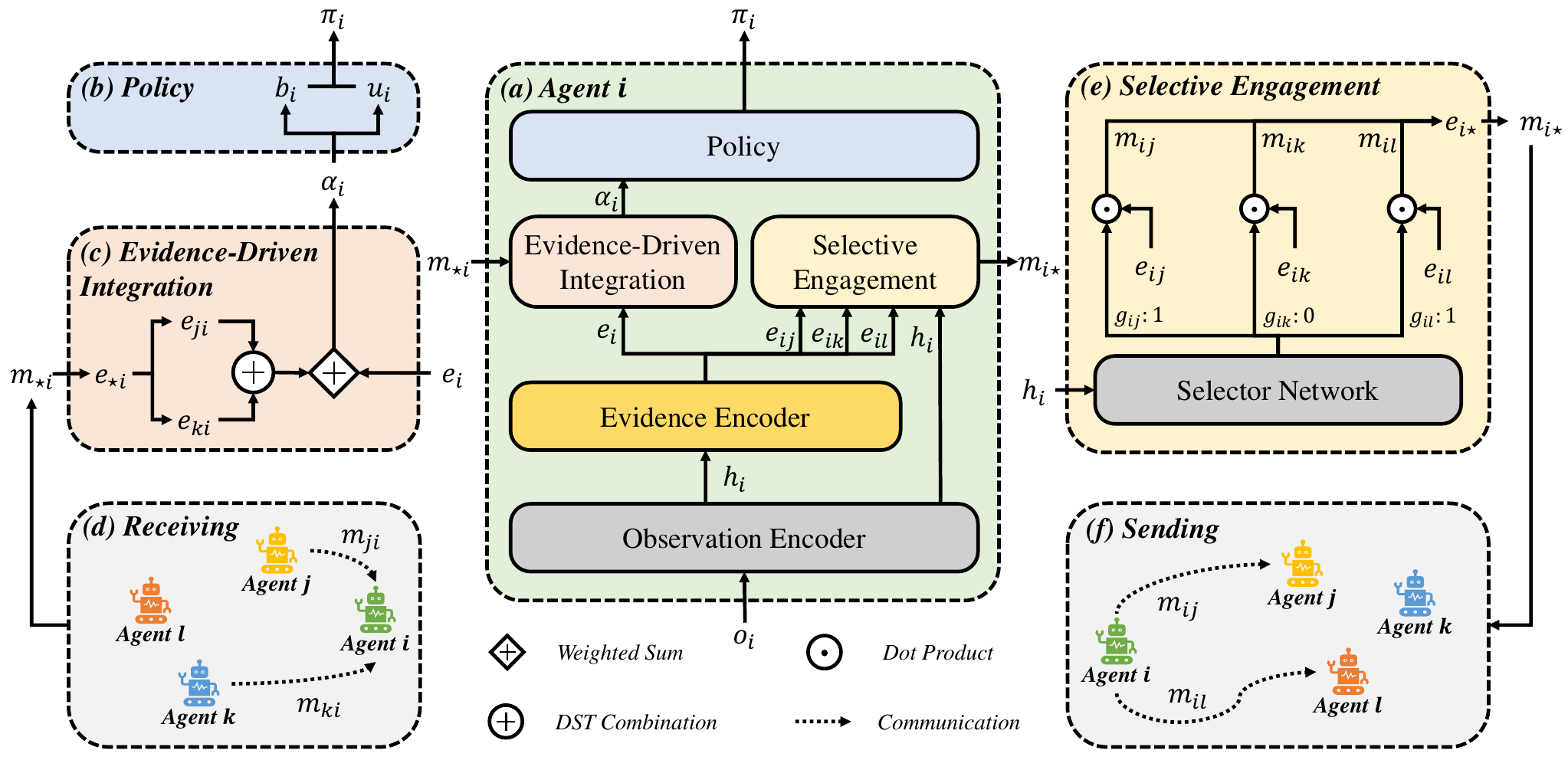}
  \caption{Framework of T2MAC.}
  \label{fig:t2mac}
\end{figure*}

\section{Background}
%需要把环境相关的且在后文出现的所有符号交代清楚
In this study, we concentrate on fully cooperative multi-agent reinforcement learning tasks characterized by partial observability while also allowing inter-agent communication. These tasks are an evolved form of Decentralized Partially Observable Markov Decision Processes (Dec-POMDPs). Their framework uses the tuple $G = (N, S, O, A, \mathbb{O}, P, R, \gamma, M)$. In this formulation: $N=(agent_1,...,agent_n)$ depicts the collective of agents. $S$ encompasses global states, offering a comprehensive environmental overview. $O$ refers to the accessible local observations. $A$ signifies a set of available actions. $\mathbb{O}$ refers to the observation function, which describes how agents perceive the environment based on the global state. $P$ acts as the transition function, illustrating environmental dynamics. $R$ is a reward function contingent on global states and joint actions. $\gamma$ represents the discount factor, $M$ delineates the set of communicable messages.
 
At each time step, agents access only local observations, which are derived from the global state through the observation function, $\mathbb{O}(o_i^t | s)$. Simultaneously, agents are equipped with the capability to share messages, denoted as $m_i^t$. These messages might encompass observations, intentions, or past experiences. Crucially, each agent can judiciously decide when to communicate, streamlining the efficiency of the communication process. %Once an agent elects to share its message, it is broadcast to the entire agent group. 
As messages are received, agents integrate their incoming information, leading to the aggregated message $c_i^t=\sum_{j \neq i} m_j^t$. This composite data then guides their localized decision-making, encapsulated by $a_i^t = \pi(o_i^t, c_i^t)$. Following this, the environment reacts to the joint action, $a=(a_1^t,...,a_n^t)$, transitioning to the subsequent state $s^{'}$. Simultaneously, this joint action then yields a shared team reward, $r=R(s,a)$. The overarching goal is to pinpoint an optimal joint policy geared towards maximizing the expected cumulative team reward, expressed as $\mathbb{E}_{s,a}[\sum_{t=0}^{\infty} \gamma^t r]$. 

%For a more comprehensive overview of preliminaries and notations, please refer to Appendix \ref{app:notations}.

\section{Methodology}
As depicted in Fig. \ref{fig:t2mac}, the distinctive characteristics of T2MAC can be highlighted in these four aspects:
\begin{itemize}
  \item T2MAC's policy is characterized as a Dirichlet distribution, facilitating the assimilation of evidence from various sources for informed and trusted decisions.
  \item The evidence encoder serves a dual purpose: extracting evidence for its own decisions and crafting tailored messages for specific teammates.
  \item Through the selective engagement, T2MAC can pinpoint optimal moments and counterparts for communication, ensuring the dissemination of only the most pertinent and reliable data.
  \item The evidence-driven integration combines incoming messages at an evidence level, refraining from treating the fusion process as a black box.
\end{itemize}
In the following sections, we will illustrate the key components of T2MAC in detail.
\subsection{Theory of Evidence}
%As clarified in \cite{TMC, Moon etal, van Amers}, using a softmax output as confidence for predictions often leads to high confidence values, even for erroneous predictions, since the largest softmax output is used for the final prediction. Therefore, we implement evidence theory as the agent policy-generating strategy that simultaneously models each action's probability and overall uncertainty of the current policy prediction.
For communication to be precise and reliable, it's essential to factor in the uncertainties intrinsic to individual decisions. To this effect, we have incorporated the theory of evidence into multi-agent communication. Within this context, evidence pertains to metrics sourced from observations supporting decision-making processes. To get a grasp on this evidence and uncertainty, we employ the Dirichlet distribution, which has proven efficacious in mitigating the overconfidence issue \cite{33,69,82}. This distribution is characterized by its concentration parameters, represented as $\alpha = [\alpha^1,...,\alpha^K]$ where $K$ is the number of actions. These parameters share an intimate relationship with uncertainty. 
%The details of Dirichlet distribution can be found in appendix. 
Building on this, we harness SL to discern the concentration parameters. SL offers a theoretical framework for extracting the probabilities (belief masses) of disparate actions and the overarching uncertainty (uncertainty mass) tied to policy-prediction challenges. Delving deeper into decision-making quandaries, SL seeks to allocate a belief mass to each action while assigning an overarching uncertainty mass to the entire scenario based on observed evidence. Consequently, all mass values remain non-negative and their cumulative value equals one:
\begin{equation}
  u_i + \sum_{k=1}^{K}b_i^k=1
\end{equation}
where $u_i\geq 0$ signifies the overall uncertainty for $agent_i$, $b_i^k\geq 0$ denotes the belief of $agent_i$ associated with the $k^{th}$ action.

Moreover, SL elegantly bridges the evidence observed by $agent_i$, denoted as $e_i=[e_i^1,, ..., e_i^K]$, with the parameters constituting the Dirichlet distribution for $agent_i$, $\alpha_i=[\alpha_i^1,, ..., \alpha_i^K]$. Here, by employing ReLU in the final layer, all evidence values are ensured to be non-negative.  
The parameter $\alpha_i^k$ is directly influenced by $e_i^k$, specifically, $\alpha_i^k=e_i^k+1$. Subsequently, the belief mass $b_i^k$ and the overarching uncertainty $u_i$ can be deduced as:
\begin{equation}
  b_i^k=\frac{e_i^k}{S_i}=\frac{\alpha_i^k-1}{S_i}\ and \ u_i=\frac{K}{S_i}
  \label{equ:bandu}
\end{equation}
where $S_i=\sum_{k=1}^K(e_i^k+1)=\sum_{k=1}^{K}\alpha_i^k$ represents the strength of the Dirichlet distribution \cite{sl}. 
Eq. \ref{equ:bandu} captures an intuitive phenomenon: the more evidence accumulated for the $k^{th}$ action, the higher the probability attributed to that action. Inversely, when there's scant evidence, the encompassing uncertainty escalates. This belief assignment can be interpreted as a form of subjective reasoning.

To enhance decision-making precision and trustworthiness, we propose leveraging evidence collected by different agents as a foundation for decision-making. Consequently, we develop an evidence encoder to deduce bespoke evidence tailored for each agent. 
At each time-step, $agent_i$ not only produces evidence $e_i$ for its own local decision but also extracts a collection of evidence - $(e_{i1},...,e_{ij},...,e_{in})$, aimed at aiding its teammates in making more reliable choices. 
Such evidence then acts as the communication medium, enabling us to generate messages tailored for specific agents. The tailored message from $agent_i$ to $agent_j$ can be denoted as $m_{ij}=e_{ij}$.

\subsection{Selective Engagement}
%这段要扣下主题提一下广播
As we've discussed earlier, broadcast communication falls short in practical applications and results in redundant information. 
The timing of information exchange and the choice of communication partners are paramount. For precise and trustworthy message exchanges, it's vital to identify truly instrumental connections from the vast web of interactions. At a holistic level, we aim to share evidence-backed data, thus enabling recipients to make informed and reliable decisions. 
To bring this vision to fruition, we meticulously quantify the strength and relevance of each communication link between agents by performing an ablative decision-making analysis. This approach primarily seeks to quantify the variability in decision uncertainty attributable to communication.
To delve deeper into the mechanics, consider the communication from $agent_i$ to $agent_j$, denoted as $m_{ij}$. This communication's value is mathematically expressed as:
\begin{equation}
  v_{ij} = u_j - \hat{u_{j}}
\end{equation}
where $u_j$ represents the decision uncertainty for recipient $agent_j$ before communication, whereas $\hat{u_{j}}$ is the uncertainty post communication.

To foster targeted and trusted communication, we develop a communication selector network. This network aids agents in determining the right moments and partners for communication, ensuring that only the most valuable and credible information is shared. We also set a constant threshold to generate binary pseudo-labels. If the deduced communication value is below the set threshold, it implies that the message received doesn't substantially benefit the recipient agent, leading the connection to be tagged as `cut', denoted mathematically as $y_{ij}=0$. However, if the communication value exceeds the threshold, it signifies the message's importance, prompting its tag to be 'retain' with $y_{ij}=1$. This systematic labeling forms the foundation for optimizing the communication selector network, with the binary cross-entropy loss steering the fine-tuning process. 

\begin{equation}
  \mathcal{L}_{BCE} = \mathbb{E}_{i,j\sim \mathbb{Z}^n}[y_{ij}\times \log(p_{ij})+(1-y_{ij})\times \log(1-p_{ij})]
  \label{equ:bce}
\end{equation}
where $\mathbb{Z}^n$ is the set of integers from 1 to $n$, $p_{ij}$ is the output of the communication selector network, representing the likelihood of $agent_i$ choosing to communicate with $agent_j$. 

\subsection{Evidence-Driven Integration}
In T2MAC, messages exchanged among agents encapsulate evidence observed from diverse perspectives. Agents can better understand the uncertain environment by adeptly integrating these messages, resulting in more sophisticated policies. 
To this end, we incorporate the DST to integrate incoming messages. This approach facilitates the combination of evidence from different sources, culminating in a degree of belief that comprehensively reflects all gathered evidence. 
%Specially, we need to 
%Then we need to combine $V$ independent sets of policy mass assignments $\{\mathcal{M}^v\}$, where $\mathcal{M}^v=\{{b_k^v}_{k=1}^K, u^v\}$, to obtain a joint mass $\mathcal{M}=\{\{b_k\}_{k=1}^{K},u\}$.
The rule of message integration for evidence is presented as:
%When combing two policy mass assignments $\mathcal{M}^1=\{\{b_k^1\}_{k=1}^{K},u^1\}$ and $\mathcal{M}^2=\{\{b_k^2\}_{k=1}^{K},u^2\}$, the joint mass, $\mathcal{M}=\{\{b_k\}_{k=1}^{K},u\}$ is calculated by the following manner:
\begin{equation}
  \mathcal{M} = \mathcal{M}_i \oplus \mathcal{M}_j
\end{equation}
where $\mathcal{M}_i=\{\{b_i^k\}_{k=1}^{K},u_i\}$ and $\mathcal{M}_j=\{\{b_j^k\}_{k=1}^{K},u_j\}$ symbolize the joint masses derived from two distinct perspectives of evidence and $\oplus$ represents DST combination. Meanwhile, $\mathcal{M}=\{\{b^k\}_{k=1}^{K},u\}$ encapsulates the consolidated joint mass, integrating evidence from both standpoints.
The more specific integration rule can be formulated as follows:
\begin{equation}
  b^k = \frac{1}{1-C}(b_i^kb_j^k+b_i^ku_j+b_j^ku_j),\ u=\frac{1}{1-C}u_iu_j
\end{equation}
where $C=\sum_{k\neq k^{'}}b_i^kb_j^{k^{'}}$ represents the degree of disagreement between the two sets of mass values. To account for this discord, DST employs the normalization factor $\frac{1}{1-C}$ to ensure a coherent integration of the evidence from both sets. 
Intuitively, when encountering evidence and beliefs from multiple sources, DST aims to merge the common elements and sidesteps conflicting beliefs through normalization factors. The integration rule ensures: 
\begin{enumerate}
  \item If both perspectives exhibit high uncertainty (with significant values of $u_i$ and $u_j$), the resultant prediction should be treated cautiously, yielding a lower confidence level (represented by a smaller value of $b^k$).
  \item Conversely, if both viewpoints possess low uncertainty (denoted by minimal values of $u_i$ and $u_j$), the resulting prediction is likely to be made with a high degree of confidence (manifesting as a larger value of $b^k$);
  \item In situations where only one viewpoint exhibits low uncertainty (meaning either $u_i$ or $u_j$ is significantly large), the final prediction predominantly relies on the more confident viewpoint.
\end{enumerate}

\begin{figure*}[t]
\centering
\includegraphics[width=1.0\textwidth]{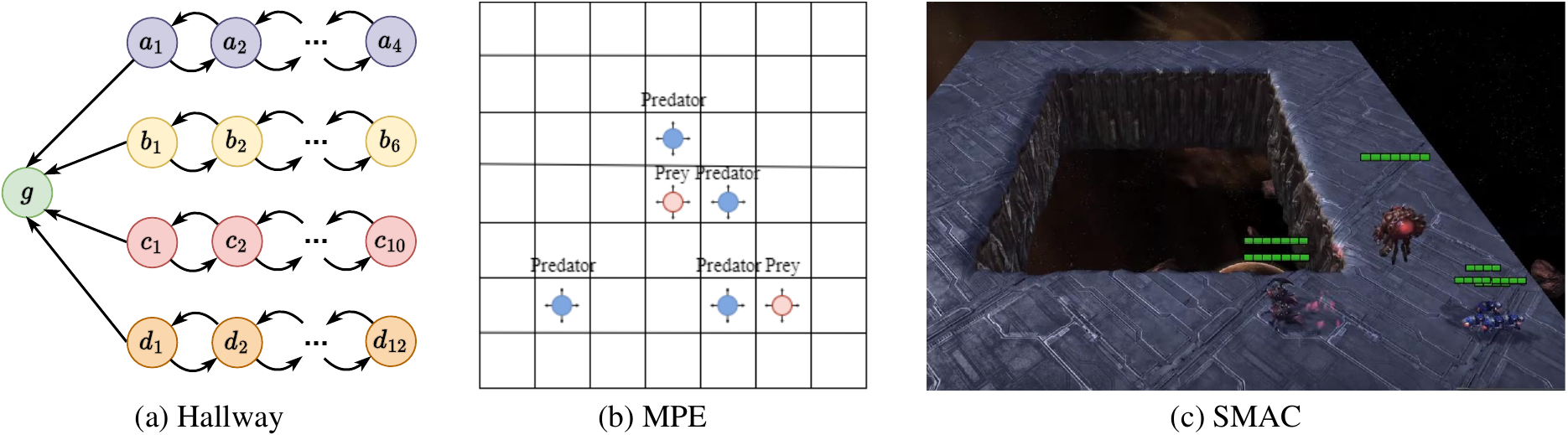}
\caption{Multiple environments considered in our experiments.}
\label{scenarios}
\end{figure*}

Upon receiving distinct messages from other agents, we derive the aforementioned mass for each perspective. Subsequently, leveraging Dempster's rule of combination, we can integrate the beliefs stemming from these varied viewpoints. More specifically, the fusion of belief and uncertainty masses across different messages is governed by the subsequent rule:
\begin{equation}
  \mathcal{M}=\mathcal{M}_1 \oplus \mathcal{M}_2 \oplus ... \mathcal{M}_n
\end{equation}
Once we have determined the joint mass $\mathcal{M}=\{\{b^k\}_{k=1}^{K},u\}$, the associated joint evidence gleaned from the messages, along with the parameters of the Dirichlet distribution, can be derived as follows:
\begin{equation}
  S=\frac{K}{u},e^k=b^k \times S\ and \ \alpha^k=e^k+1
\end{equation}
Leveraging DST, we attain an efficient and theoretically-founded method for message integration. This method skillfully amalgamates messages and simultaneously addresses enduring intrinsic policy uncertainties. Importantly, the fusion of information isn't treated as a black box, given that the combination rules of DST lack learnable parameters. Furthermore, DST offers a more comprehensible and theoretical perspective on the message integration process. 

Following the assimilation of incoming messages and the acquisition of integrated evidence, each agent makes a local decision influenced by both its observed and received evidence. For $agent_i$, this procedure is represented as:
\begin{equation}
  a_i^t = \pi_i(\hat{e_i})
\end{equation}
where $\hat{e_i}$ symbolizes the evidence post-integration for $agent_i$ at time-step $t$. 
For details of the communication process and the training paradigm of T2MAC, please refer to pseudo-code  provided in  this section. 
\begin{algorithm}[t]
\begin{algorithmic}
    \STATE Initialize replay buffer D
    \STATE Initialize the Observation encoder, Evidence encoder, Selective Engagement and  Q network with random parameters
    %\STATE Initialize $\theta$, the parameters for Policy $\pi$; $\phi$, the parameters for Selective Engagement
    \STATE Set learning rate $\alpha$ and max training episode $E$
    \FOR {episode in $1,...,E$}
        \FOR{each agent $i$ }
            \STATE \textbf{Sending Phase:} Encode the hidden feature $h_i^t$ from observation $o_i^t$
            \STATE Encode evidence $e_i^t$ for local decision
            \STATE Encode evidence and generate tailored messages for specific teammates $(e_{i1},...,e_{ij},...,e_{in})$
            \STATE Select ideal communication partners using communication selector network
            \STATE \textbf{Receiving Phase:} Combing received messages $m_{\star i}^t$ from other agents by DST combine
            %\STATE Combing $m_{\star i}$ and $e_i^t$ by weighted sum
            \STATE Select action $a_i^t$ by combined evidence
            \STATE Compute the importance for each communication link and generating labels $y_{ij}$ for communication selector network
        \ENDFOR
        \STATE Store the trajectory in replay buffer D
        \STATE Sample a minibatch of trajectories from D
        \STATE Update observation encoder, evidence encoder and policy network using MARL loss function %denoted in \cite{qmix}
        \STATE Update Selective Engagement by Equation \ref{equ:bce}
    \ENDFOR
\end{algorithmic}
\caption{$\text{T2MAC}$}
\label{alg:T2MACfull}
\end{algorithm}

%The pseudo-code for T2MAC can be found in the Appendix \ref{app:pesudocode}.
%Based on the above combination rule, we can obtain the estimated multi-view joint evidence $e_m$ and combine it with the evidence $e_i$ produced by the agent $i$ itself:
%\begin{equation}
%  e_i^{combined}=e_i + w\times e_m
%\end{equation}
%where $w$ is the sum weight.

\begin{figure*}[t]
   \centering
   \includegraphics[width=1.0\textwidth]{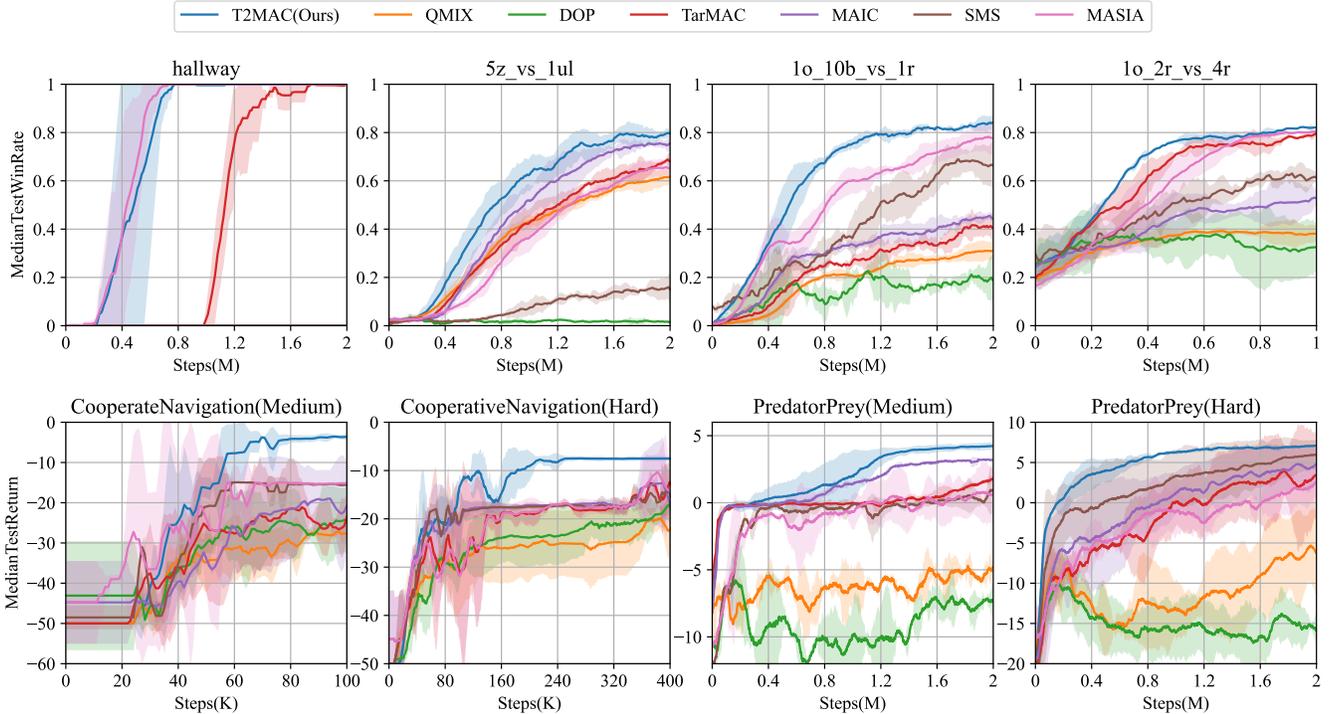}
   \caption{Performance on multiple benchmarks.}
   \label{performance}
 \end{figure*}

\section{Experiments}
In this section, we carefully design experiments to address three pivotal questions: (1) How does T2MAC's performance measure against top-tier communication methods? (2) What characterizes T2MAC's communication efficiency? (3) Can T2MAC scale across various tasks and seamlessly integrate with multiple baselines? 

\subsection{Setup}
As illustrated in Fig. \ref{scenarios}, we extensively evaluate T2MAC across three notable cooperative multi-agent tasks. Beginning with Hallway \cite{ndq}, this environment is relatively direct, built around multiple Markov chains. Here, agents start at random positions within different chains and aim to reach the goal state simultaneously under partial observability. To escalate the complexity, we augment the number of agents and the length of the Markov chains, leading to a substantial increase in the exploration space. 
On the other hand, MPE \cite{maddpg} is a vital MARL benchmark set in a 2D grid. We focus on the Cooperative Navigation (CN) and Predator Prey (PP) scenarios. In CN, the task for agents is to navigate to different landmarks, whereas, in PP, their objective is to capture unpredictably moving prey. To introduce varying difficulty levels, we employ different grid sizes for both scenarios. The Cooperative Navigation: Medium scenario is set on a $7\times7$ grid, while the Cooperative Navigation: Hard occupies a $9\times 9$ grid. The Predator Prey: Medium scenario is set on a $5\times5$ grid, while the Predator Prey: Hard occupies a $7\times 7$ grid. SMAC \cite{smac} is derived from the well-known real-time strategy game StarCraft \uppercase\expandafter{\romannumeral2}. It delves into micromanagement challenges where each unit is steered by an independent agent making decisions under partial observability. To emphasize the importance of communication, we adopt the setup from \cite{ndq}, which not only restricts the agents' sight range %, reducing it drastically from 9 to just 2 
but also throws them into intricate maps, characterized either by their labyrinthine terrains or the unpredictable spawning dynamics of units. 
For comparative analysis, we draw from a diverse set of baselines. This includes non-communication paradigms like the leading MARL methods QMIX \cite{qmix} and DOP \cite{dop}. Meanwhile, our baselines include contemporary state-of-the-art communication methods, such as TarMAC \cite{tarmac}, MAIC \cite{maic}, SMS \cite{sms}, and MASIA \cite{masia}.

In conclusion, our experimental design integrates a medley of challenging tasks and robust baselines, establishing a solid foundation for evaluation. 
Our overarching goal with this varied selection is to place T2MAC in diverse scenarios and test its adaptability, scalability, and overall performance. %The results, framed in terms of median performance metrics bolstered by a 95\% confidence interval, should provide a clear indication of where T2MAC stands amidst current methodologies.
To ensure transparency and reproducibility,  the intricate details of our method's architecture  and our hyperparameter choices are extensively detailed in Table \ref{tab:netarch}.

\begin{table}[t]
\centering
\begin{tabular}{ll}
  \toprule
  \textbf{Module}  & Architecture \\
  \midrule
  \multirow{4}*{Obs Encoder}      & Linear(obs\_dim, 64)\\
                                  & Linear(64, 64)\\
                                  & Linear(64, 64)\\
                                  & RNN(64, 64) \\\hline
  Evidence Encoder                & n*Linear(64, K)\\\hline
  Selector Network & Linear(64, n) \\
  \bottomrule
\end{tabular}
\caption{Hyperparameters of T2MAC}
\label{tab:netarch}
\end{table}

%\begin{table*}[t]
% \caption{The detailed information of selected SMAC scenarios}
% \label{tab1}
%% \begin{tabular}{cccc}\toprule
 %  \textit{} & \textit{Difficulty} & \textit{Allied units} & \textit{Enemy units} \\ \midrule
 %  5m\_vs\_6m & Hard & 5 Marines & 6 Marines \\
 %%  27m\_vs\_30m & Super Hard & 27 Marines & 30 Marines \\
 % 3s5z\_vs\_3s6z & Super Hard & 3 Stalkers and 5 Zealots & 3 Stalkers and 6 Zealots  \\
 % 6h\_vs\_8z & Super Hard & 6 Hydralisks & 8 Zealots \\
 % corridor & Super Hard & 6 Zealots & 24 Zerglings\\
 % MMM2 & Super Hard & 1 Medivac, 2 Marauders and 7 Marines & 1 Medivac, 3 Marauders and 8 Marines \\ \bottomrule
 %\end{tabular}
%\end{table*}

\subsection{Results}

\subsubsection{Performance}
We begin our evaluation by comparing the learning curves of T2MAC with various baselines across various environments to test its overarching performance. 
As illustrated in Fig. \ref{performance}, T2MAC emerges superior in almost all environments, highlighting its robust performance. 
In Hallway, as the difficulty intensifies, many baselines falter, unable to adapt effectively. Among them, only MASIA stands out, delivering commendable results, primarily due to its ability to assist agents in reconstructing global information. Intriguingly, our T2MAC works even under such demanding conditions, achieving performance on par with MASIA. This might be largely attributed to its adeptness at sharing and integrating relevant evidence.
In SMAC, T2MAC delivers consistent and impressive performance across all three maps. However, when looking at all scenarios in their entirety, other methods exhibit signs of instability. For instance, SMS struggles to adapt in the $5z\_vs\_1ul$, while TarMAC fails in the $1o\_10b\_vs\_1r$. Such observations accentuate, to some extent, the broad applicability and robustness inherent to T2MAC. 
In CN and PP, T2MAC maintains its sustained sample efficiency. Upon reaching a convergence, its performance remains fiercely competitive. 
Furthermore, an interesting observation is that all methods incorporating communication significantly outperform those that don't. This emphasizes that our chosen environments and scenarios intrinsically demand proficient communication. Such an outcome not only underscores the importance of communication in these contexts but also validates the aptness of our experimental setup in benchmarking communication methods.

\begin{table}[t]
\centering
\begin{tabular}{l|lll}
  \toprule
  \makecell[c]{Methods} & \thead{Performance \\ Improvement} & \thead{Comm \\ Rate} & \thead{Comm \\ Efficiency} \\\hline
  \makecell[c]{TarMAC} &   \makecell[c]{17.0\%}   &   \makecell[c]{ 100.0\%}    &   \makecell[c]{17.0\%}\\
  \makecell[c]{MAIC}&   \makecell[c]{12.3\%}  &  \makecell[c]{100.0\%}     &   \makecell[c]{12.3\%} \\
  \makecell[c]{SMS}&  \makecell[c]{27.9\%}  &\makecell[c]{66.7\%}   &     \makecell[c]{ 41.8\%}    \\
  \makecell[c]{MASIA} &  \makecell[c]{30.2\%} & \makecell[c]{100.0\%}  & \makecell[c]{30.2\%}  \\\hline
  \makecell[c]{T2MAC(Ours)} & \makecell[c]{\textbf{37.2\%}}& \makecell[c]{\textbf{56.0\%}} & \makecell[c]{\textbf{66.4\%}} \\
  \bottomrule
\end{tabular}
\caption{Communication Efficiency}
\label{table:efficiency}
\end{table}

\subsubsection{Efficiency}
In addition to analyzing the overarching performance, we also focus on understanding communication efficiency. In many real-world situations, communication resources—like bandwidth and transmission channels—are inherently scarce. Overloading these resources doesn't always yield proportional benefits in performance. To quantify this efficiency, we calculate the performance improvement attributable to communication and then normalize this by the communication rate. Here, communication rate denotes how frequently communication occurs throughout the learning process. To gauge performance improvement, we introduce a communication-free variant for each communication method. This allows us to make a side-by-side comparison to effectively highlight the tangible advantages offered by each method. Specifically, for SMS, this communication-free baseline is DOP, while for the others, it's QMIX. We've carried out this analytical assessment predominantly in the most challenging environment, SMAC. As shown in Table. \ref{table:efficiency}, T2MAC consistently outperforms baselines in terms of both improvement, communication rate, and communication efficiency. Such results underscore the capability of T2MAC to process communication dynamics, including when to communicate, with whom, and how to trade-off between performance and efficiency.

\subsubsection{Generality} 
\begin{figure}[t]
  \centering
  \includegraphics[width=0.4\textwidth]{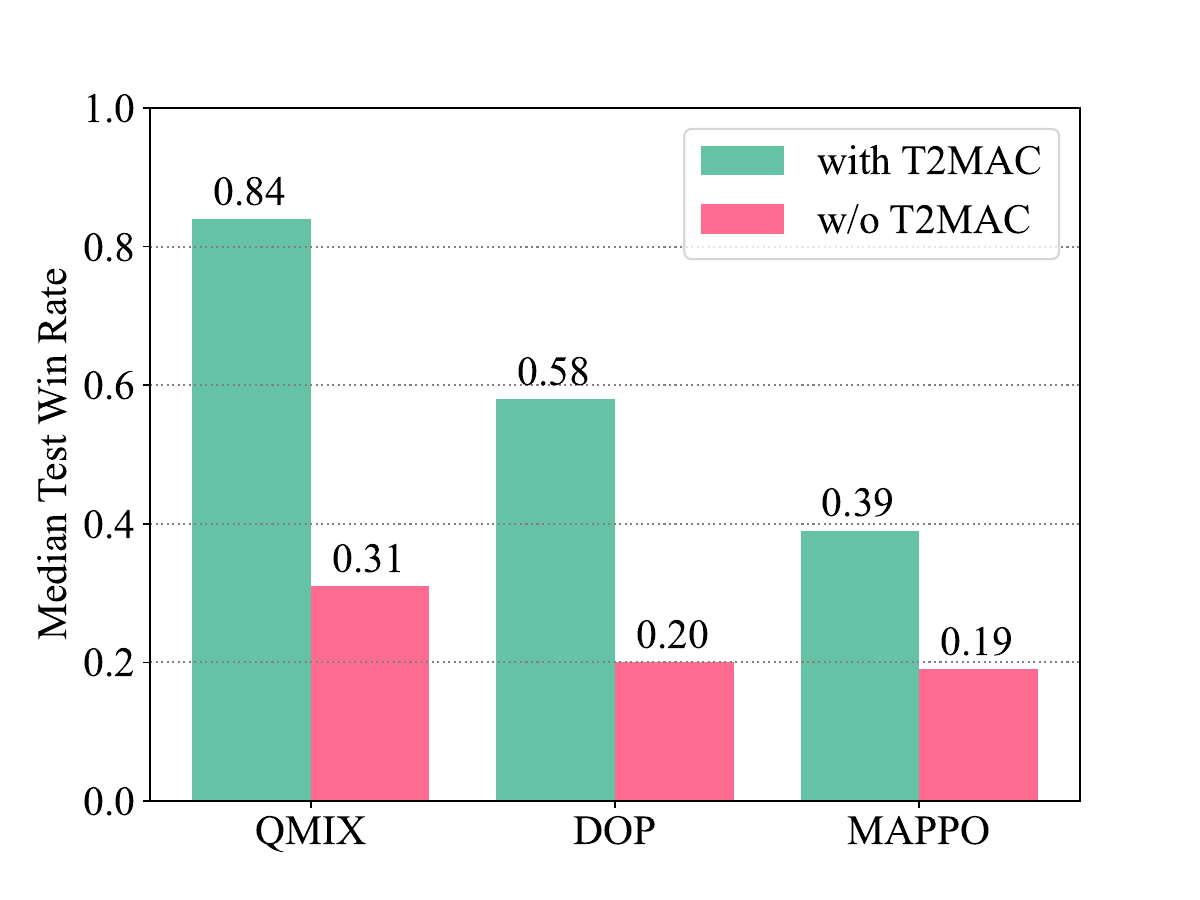}
  \caption{Generality.}
  \label{generality}
\end{figure}

Our prior experiments have demonstrated the robustness of T2MAC across diverse environments, scenarios of varying complexities, and different scales. To further evaluate the generality of T2MAC, we apply it across a wide range of established MARL baselines, including QMIX, DOP, and MAPPO. The test win rate for the scenario $1o\_10b\_vs\_1r$ is illustrated in Fig. \ref{generality}. 
Notably, across all these baselines, T2MAC consistently achieves superior performance, often by a notable margin. This positive performance improvement demonstrates the broad applicability and potency of T2MAC in the realm of MARL.
\subsubsection{Ablation}
\begin{figure}[t]
  \centering
  \includegraphics[width=0.35\textwidth]{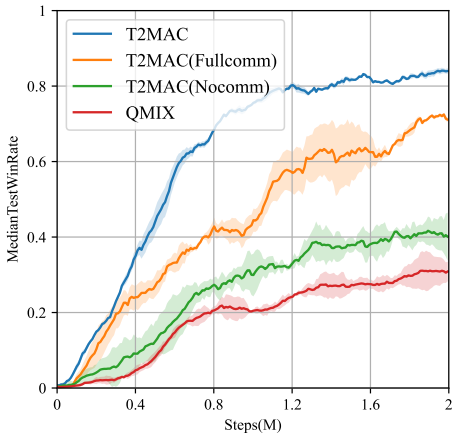}
  \caption{Ablation for trusted communication and selective engagement.}
  \label{ablation}
\end{figure}

To better understand the impact of each component within T2MAC, we perform an ablation study on the scenario $1o\_10b\_vs\_1r$. Here's a breakdown of the configurations evaluated: \textbf{T2MAC}: This refers to the complete method proposed in our work. \textbf{QMIX}: This serves as our baseline for comparison, representing the core functionality without the enhancements introduced in T2MAC. \textbf{T2MAC(Fullcomm)}: This is a variant of T2MAC that does not incorporate selective engagement. Here, communication occurs continuously amongst agents without deciding when or with whom to communicate. \textbf{T2MAC(Nocomm)}: This is a more stripped-down version of T2MAC, excluding both selective engagement and evidence-driven integration. Essentially, it's a version of T2MAC where communication is completely omitted, but the Dirichlet Distribution remains in the Q-value network.
As illustrated in Fig. \ref{ablation}, the results demonstrate the contributions of each component: \textbf{From QMIX to T2MAC(Nocomm): }The shift from Categorical distribution to Dirichlet distribution makes sense. The Dirichlet distribution's advantage might stem from its ability to model second-order probabilities, introducing an additional layer of decision-making uncertainty which potentially enhances learning and adaptation. \textbf{From T2MAC(Nocomm) and T2MAC(Fullcomm)}: The sizable performance gap between these two underscores the significance of evidence-driven information exchange and integration. This sheds light on the efficacy of trust-based communication, where agents not only share but also assess the reliability of information before acting upon it. \textbf{From T2MAC(Fullcomm) to T2MAC}: The contrast in performance between these two configurations underlines the importance of targeted communication. Instead of a blanket communication strategy, selective engagement, whereby agents communicate at strategic junctures with specific partners, can enhance the overall efficiency and performance of the system. 

\begin{figure}[t]
  \centering
  \includegraphics[width=0.35\textwidth]{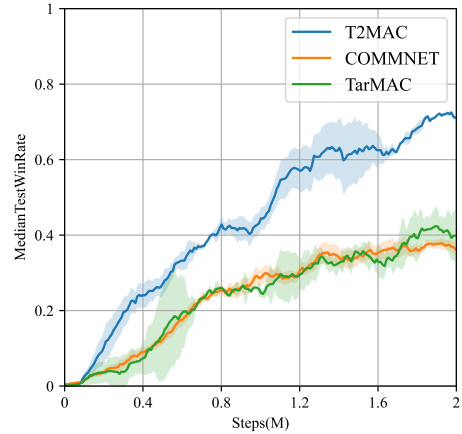}
  \caption{Ablation for evidence-driven integration.}
  \label{ablation2}
\end{figure}

Furthermore, to provide a clear ablation analysis for evidence-driven integration, we have conducted additional comparisons in the $1o\_10b\_vs\_1r$ scenario with a summation-based integration method (COMMNET\cite{commnet111}) and a black-box method (TarMAC\cite{tarmac}). As shown in Fig. \ref{ablation2}, the results demonstrate that the evidence-driven integration proposed by T2MAC has a clear advantage, confirming its effectiveness.

\section{Conclusions}
In this work, we tackle the intricacies inherent in multi-agent communication. Previous works focus on broadcast communication and treat the fusion of information as a block box, which inevitably diminishes communication efficiency. To this end, we present the T2MAC framework. This novel approach empowers agents with the capacity to craft messages specifically tailored for distinct agents. Beyond mere message customization, T2MAC strategically chooses the best timings and relies on trusted partners for communication, ensuring an efficient integration of incoming messages and facilitating trusted decision-making. Rooted in solid theoretical principles, this approach stands out for its efficiency. Furthermore, to substantiate our claims, we conduct comprehensive experiments across multiple benchmarks, the results of which underscore the effectiveness, efficiency, and adaptability of the T2MAC.

%%%%%%%%%%%%%%%%%%%%%%%%%%%%%%%%%%%%%%%%%%%%%%%%%%%%%%%%%%%%%%%%%%%%%%%%%%%%
%%%%%%%%%%%%%%%%%%%%%%%%%%%%%%%%%%%%%%%%%%%%%%%%%%%%%%%%%%%%%%%%%%%%%%%%%%%%
%%%%%%%%%%%%%%%%%%%%%%%%%%%%%%%%%%%%%%%%%%%%%%%%%%%%%%%%%%%%%%%%%%%%%%%%%%%%
%%%%%%%%%%%%%%%%%%%%%%%%%%%%%%%%%%%%%%%%%%%%%%%%%%%%%%%%%%%%%%%%%%%%%%%%%%%%
\section*{Acknowledgements}
The authors would like to thank the editors and reviewers for their valuable comments. This work is supported by the Youth Innovation Promotion Association CAS, No. 2021106, the China Postdoctoral Science Foundation, No. 2023M743639, the 2022 Special Research Assistant Grant project, No. E3YD5901, and the CAS Project for Young Scientists in Basic Research, Grant No. YSBR-040.

% The authors would like to thank the editors and reviewers for their valuable comments. This work is supported by 2022 Special Research Assistant Grant project, No. E3YD5901, the China Postdoctoral Science Foundation, No. 2023M743639, the Youth Innovation Promotion Association CAS, No. 2021106, and the Fundamental Research Program, Grant No. JCKY2022130C020.
%%%%%%%%%%%%%%%%%%%%%%%%%%%%%%%%%%%%%%%%%%%%%%%%%%%%%%%%%%%%%%%%%%%%%%%%%%%%
%%%%%%%%%%%%%%%%%%%%%%%%%%%%%%%%%%%%%%%%%%%%%%%%%%%%%%%%%%%%%%%%%%%%%%%%%%%%
%%%%%%%%%%%%%%%%%%%%%%%%%%%%%%%%%%%%%%%%%%%%%%%%%%%%%%%%%%%%%%%%%%%%%%%%%%%%
%%%%%%%%%%%%%%%%%%%%%%%%%%%%%%%%%%%%%%%%%%%%%%%%%%%%%%%%%%%%%%%%%%%%%%%%%%%%

\bibliography{aaai24}

\end{document}